# Toward arbitrary spin-orbit flat optics via structured geometric phase gratings

Chun-Yu Li[1,†], Si-Jia Liu[2,†], Bing-Shi Yu[1], Hai-Jun Wu[1], Carmelo Rosales-Guzmán[1,3], Yijie Shen[4], Peng Chen[2*], Zhi-Han Zhu[1*] and Yan-Qing Lu[2*]

[1] Wang Da-Heng Center, Heilongjiang Key Laboratory of Quantum Control, Harbin University of Science and Technology, Harbin 150080, China
[2] National Laboratory of Solid State Microstructures, Key Laboratory of Intelligent Optical Sensing and Manipulation, and College of Engineering and Applied Sciences, Nanjing University, Nanjing 210093, China
[3] Centro de Investigaciones en Óptica, A.C., Loma del Bosque 115, Colonia Lomas del Campestre, 37150 León, Gto., Mexico
[4] Optoelectronics Research Centre & Centre for Photonic Metamaterials, University of Southampton, Southampton SO17 1BJ, United Kingdom
* e-mail: chenpeng@nju.edu.cn, zhuzhihan@hrbust.edu.cn, and yqlu@nju.edu.cn
† These authors contributed equally.

**Abstract**: Reciprocal spin-orbit coupling (SOC) via geometric phase with flat optics provides a promising platform for shaping and controlling paraxial structured light. Current devices, from the pioneering q-plates to the recent J-plates, provide only spin-dependent wavefront modulation without amplitude control. However, achieving control over all the spatial dimensions of paraxial SOC states requires spin-dependent control of corresponding complex amplitude, which remains challenging for flat optics. Here, to address this issue, we present a new type of flat-optics elements termed structured geometric phase gratings that is capable of conjugated complex-amplitude control for orthogonal input circular polarizations. By using a microstructured liquid crystal photoalignment technique, we engineered a series of flat-optics elements and experimentally showed their excellent precision in arbitrary SOC control. This principle unlocks the full-field control of paraxial structured light via flat optics, providing a promising way to develop an information exchange and processing units for general photonic SOC states, as well as extra-/intracavity mode convertors for high-precision laser beam shaping.

## Introduction

Recent advances in structured vectorial paraxial beams with tunable spatially-varying amplitude, phase and state of polarization (SoP) are continually revealing a variety of exotic structured photonic states and unexpected phenomena [1-4]. These vector paraxial modes were discovered in the early years of laser physics but became an active topic only in this century [5]. The mechanism behind the exotic beam structure of vector modes has become widely known only in the last two decades, after people became aware of paraxial orbital angular momentum (OAM) [6,7]. More precisely, vector modes a nonseparable superposition between the SoP and spatial structure of light, which can also be interpreted as spin-orbit coupling (SOC) in light beams, and the modes are therefore also known as paraxial SOC states, especially in the quantum context [8-10]. Crucially, this mechanism has provided the basis on how to shape and control the SOC states — i.e., spin-dependent spatial light modulation. Along this line, there are two feasible approaches: the first one relies on the use of a traditional polarization interferometer, very often comprising a digital spatial light modulator [11,12], while the other is to exploit the geometric phase, through photonic SOC devices based on delicate flat optics [13]. The former can fully control the spatiotemporal structure of light in a dynamic and flexible way but has a complex, bulky and inefficient configuration. The latter, owing to its low cost, integrability and versatility, is gaining increasing attention and popularity in a rapidly developing field [14,15].

The geometric phase referred here originates from the slow transformation of the SoP, also known as the Pancharatnam–Berry phase, and its retardance depends on the geometric path on the Poincaré sphere [16-18]. This principle provides a new paradigm for building wavefront-shaping platforms, i.e., exploiting the 2D geometric phase resulting from a spatially varied SoP transformation. Specifically, imagine if we pass a beam through a flat element with spatial-variant birefringence, such as microstructured liquid crystals (LCs) or dielectric metasurfaces [14,15], beyond enabling a point-by-point SoP transformation in the transverse plane, a wavefront modulation with SoP switchable behaviour will be achieved. Notably, current SOC devices, from the original LC q-plates to recent metasurface J-plates [19,20], provide only spatial light modulation of the phase with SoP-switchable behaviour. The absence of amplitude control greatly limits the range of SOC achievable states. To illustrate this, we take cylindrical vector (CV) polarized modes (also called vector vortex beams) as an example, which have been considered of paramount importance in applications, such as, high-dimensional communication, remote sensing, microscopy, to mention a few [21-28]. Their modal constitution exhibits 'typical' SOC states, i.e., a superposition of two conjugate (opposite topological charge) OAM modes with orthogonal circular polarization. Remarkably, CV modes obtained via q-plates are not eigenmodes of free space and therefore, upon propagation they give rise to the so-called hyper-Geometric Gauss (HyGG) beams [29]. In contrast to propagation invariant Laguerre–Gauss (LG) modes, their pattern exhibits ripple-like motion upon propagation owing to their undefined amplitude structure [30]. This issue creates difficulties in SOC state manipulation and transmission, but more importantly, it limits the exploitation of all the spatial dimensions of structured light. Without amplitude control, it is impossible to access the radial degree-of-freedom of LG and associated CV modes [31-35], let alone to control more general SOC states in other paraxial coordinates [36-38]. That is why q-plate-like flat elements have always been absent from experimental studies involving full spatial dimensions of structured light, and instead, the traditional optical scheme seems to be the only choice of scientists [39-42].

To control a paraxial SOC state in all its spatial degrees of freedom, spin-dependent complex amplitude modulation provides an essential alternative, but up to now it has remained elusive with flat optics. In this work we fill this gap, by putting forward a new type of geometric-phase element termed Structured Geometric Phase Grating (SGPG), featuring a spatially-varying grating cycle, depth and orientation. Importantly, this SOC device can structure arbitrarily the complex amplitude of input beams in the $\pm 1^{\text{st}}$-order of diffractions with a conjugate manner in response to left- and right-circular polarized components, respectively. Such a crucial advance, compared with the present geometric phase elements, unlocks the control of paraxial structured light in all spatial dimensions, and paves the way towards arbitrary SOC conversion via flat optics. To demonstrate this principle, we engineered a series of LC flat elements using a photoalignment technique [15] and experimentally showed their excellent performance in terms of arbitrary SOC control of structured light.

## Concept and Principle

**Spin-orbit coupled states** — The term SOC state utilized here refers to the most common family of vectorial paraxial modes constructed by conjugate angular momentum components [43] and can be expressed as a nonseparable superposition of orthogonal circular SoPs $\hat{e}_\pm$ (associated to spin angular momentum) and conjugate spatial modes $\psi_\pm(r, z)$ carrying opposite OAM

$$\Psi_{\text{soc}}(r, z) = \sqrt{a}\psi_+(r,z)\hat{e}_+ + e^{i\theta}\sqrt{1-a}\psi_-(r,z)\hat{e}_-, \quad (1)$$

where $r$ denotes transverse coordinates, $a \in [0,1]$ is a weighting coefficient that controls the degree of nonseparability (which is maximum for $a = 0.5$), and $\theta$ is an intermodal phase. Note that the two OAM-carrying modes $\psi_\pm(r, z)$ form a conjugate pair and thus can be expressed as

$$\psi_\pm(r, z) = u(r, z)\exp[\pm i v(r, z)], \quad (2)$$

where $u(r, z)$ and $\pm v(r, z)$ represent the identical spatial amplitude and conjugate wavefronts of the pair, respectively. All possible vector modes shown in Eq. (1) form a tensor parameter space with respect to spin and orbital angular momenta and can be visualized as the surface of a spin-orbit hybrid unit sphere commonly called higher-order Poincaré sphere [44], although less rigorous [45]. Here we call it SOC modal sphere in the following contents. In particular, the SOC state (1) becomes the most common CV mode, as $\psi_\pm(r, z)$ are OAM eigenmodes carrying well-defined OAMs, i.e., $\pm \ell \hbar$ per photon ($\ell$ is an integer), such as a pair of conjugate LG, Bessel or HyGG modes.

Note that state (1) has the same SU(2) algebraic structure as a scalar SoP on the 'classical' Poincaré sphere. This indicates that a reciprocal SOC device, if achievable, can realize the interconversion between an arbitrary SOC mode and its corresponding scalar SoP

$$(\sqrt{a}\hat{e}_+ + e^{i\theta}\sqrt{1-a}\hat{e}_-)\psi_0(r, z) \leftrightarrow \Psi_{\text{soc}}(r, z), \quad (3)$$

where $\psi_0(r, z)$ denotes the spatial complex amplitude of the scalar beam and is commonly considered the TEM$_{00}$ mode. In other words, the supposed device can map a SoP on the Poincaré sphere into the same position on an arbitrary desired SOC modal sphere and vice versa, as the example shown in Fig. 1(a). A popular example is the q-plate, used to interconvert a given scalar SoP and its CV-polarized counterpart. The mechanism of this interconversion is realized by imprinting conjugate spiral wavefronts $\exp(\pm i \ell \varphi)$ onto $\hat{e}_\pm$ polarization components of the input beam. This widely used spin-dependent phase-only modulation [13, 15, 20], although belonging to unitary transformation, fails to define the amplitude structure of $\psi_\pm(r, z)$. Namely, the implementation of Eq. (3) for general SOC states relies on a reciprocal complex-amplitude transformation $\psi_\pm(r, z) \leftrightarrow \psi_0(r, z)$ according to the input SoP. In particular, this arbitrary SOC conversion can work as a controlled gate in quantum computing system built by photonic SOC states [46].

**Structured geometric phase gratings** — How to build a flat SOC device for structuring arbitrary paraxial vector modes? The task requires that the device, beyond conjugately shaping wavefronts of $\hat{e}_\pm$ components, can also structure the amplitude profile directly and preferably without the assistance of other operations, such as polarization filtering. Moreover, the intramodal phase $e^{i\theta}$ should remain unchanged in the conversion. That is, our main focus is to achieve spin-dependent complex amplitude modulation with phase-locking ability.

To achieve this goal, we introduce a new type of geometric-phase element termed the Structured Geometric Phase Grating (SGPG) that has spatially-variant grating cycle, depth and orientation. When guiding a TEM$_{00}$ or $\Psi_{\text{soc}}(r, z)$ mode passing this flat element, the designed mode conjugates $\psi_\pm(r, z)$ can be generated or measured at the $\pm 1^{\text{st}}$ diffraction orders with opposite circular SoPs. The two diffraction orders can be recombined coherently using a conventional polarizing grating (PG); in this way, the reciprocal SOC device for arbitrary vector modes shown in Eq. (3) is realized. Both microstructured LCs and dielectric metasurfaces are good candidates to fabricate SGPGs. Here, we used nematic LC and the photoalignment technique to fabricate all the elements by virtue of their high efficiency, low cost, high reliability and electro-tunability [47-50]; see Methods for details. These LC elements can be regarded as microstructured half-wave plates with spatially-variant director orientation $\alpha(r)$, giving a 2D SoP transformation and associated wavefront modulation $|\hat{e}_\pm\rangle \to \exp[\pm i2\alpha(\boldsymbol{r})]|\hat{e}_\mp\rangle$ to circular-polarized components of incident light.

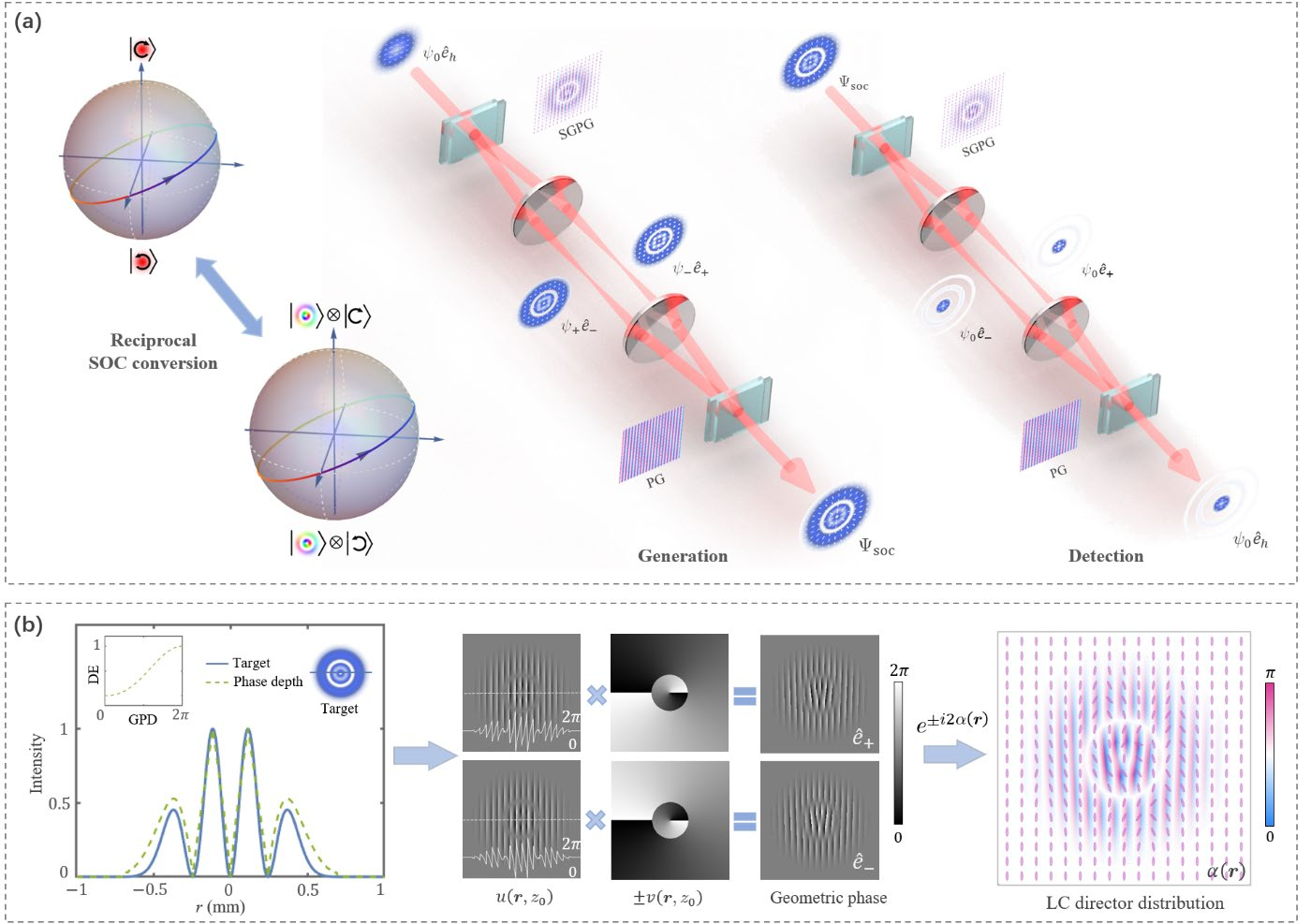

**Figure 1. (a)** Schematic of the reciprocal SOC device based on structured geometric phase gratings (SGPGs) designed for controlling the 'example' SOC state composed of $|LG_{+1,1}, \hat{e}_+\rangle$. **(b)** Inverse design of the SGPG for generating and measuring $|LG_{+1,1}, \hat{e}_+\rangle$ in $\pm 1^{st}$-order diffractions. The left panel shows the target intensity profile to be structured, diffraction efficiency (DE) and the corresponding grating phase depth (GPD) distribution, the middle panel shows the constructed spin-switchable geometric phase, and the right panel shows the LC-director orientations of the designed SGPG.

To illustrate the principle specifically, as shown in Fig. 1, we choose vector modes constructed by SOC bases $|LG_{\pm 1,1}, \hat{e}_\pm\rangle$ as an example to show the inverse design of the geometric-phase elements. That is, in the example, the desired mode conjugates $\psi_\pm(\boldsymbol{r}, z)$ were assumed as a dual-ring donut mode pair $LG_{\pm 1,1}$. First, to directly structure the spatial amplitude via geometric phase, we exploit the relation of the grating phase depth ($D$) to diffractive efficiency (or probability) to control amplitude profiles of light (or photons) in $\pm 1^{st}$-order diffractions, given by

$$D(\boldsymbol{r}) = 2\pi\{1 - \operatorname{sinc}^{-1}[u(\boldsymbol{r}, z_0)]\}, \qquad (4)$$

where $D \in [0, 2\pi]$ and $u(\boldsymbol{r}, z_0)$ denote the amplitude profile (realized via diffractive efficiency control) of desired spatial modes at the $z = 0$ plane. Specifically, in the left panel of Fig. 1(b), the green dashed line shows a calculated grating phase depth $D(\boldsymbol{r})$ with the blue solid curve as the desired diffractive efficiency (or the beam profile to be shaped) $u^2(\boldsymbol{r})$. Based on this, by integrating with the spin-dependent wavefronts $\pm v(\boldsymbol{r}, z_0)\hat{e}_\pm$, a pair of geometric phase conjugates used to structure the example mode pairs $LG_{\pm 1,1}$ is obtained, as shown in the middle panel. At last, using the relation between the geometric phase and LC-director orientation $|\hat{e}_\pm\rangle \to \exp[\pm i2\alpha(\boldsymbol{r})]|\hat{e}_\mp\rangle$, we obtain the LC-director distribution to fabricate the desired SGPG, as shown in the right panel. This novel geometric-phase element can exactly structure the example SOC bases $|LG_{\pm 1,1}, \hat{e}_\pm\rangle$ in $\pm 1^{st}$-order diffractions. On this basis, in combination with a 4*f*-PG polarizing beam combiner, we build a reciprocal SOC device that can map arbitrary scalar SoPs into their vector counterpart on the SOC modal sphere defined by $|\text{LG}_{\pm 1,1}, \hat{e}_{\pm}\rangle$ and vice versa. In addition, there are two noteworthy points. First, incident beams in the generation process are usually in the fundamental Gaussian mode; thus, the corresponding design should contain Gaussian amplitude amendments. Second, the principle of spatial mode detection used here, as well as in other relevant works, is not a strict projective measurement but a spatial autocorrelation [51-53]. More details about the two points are provided in the Supplementary Materials.

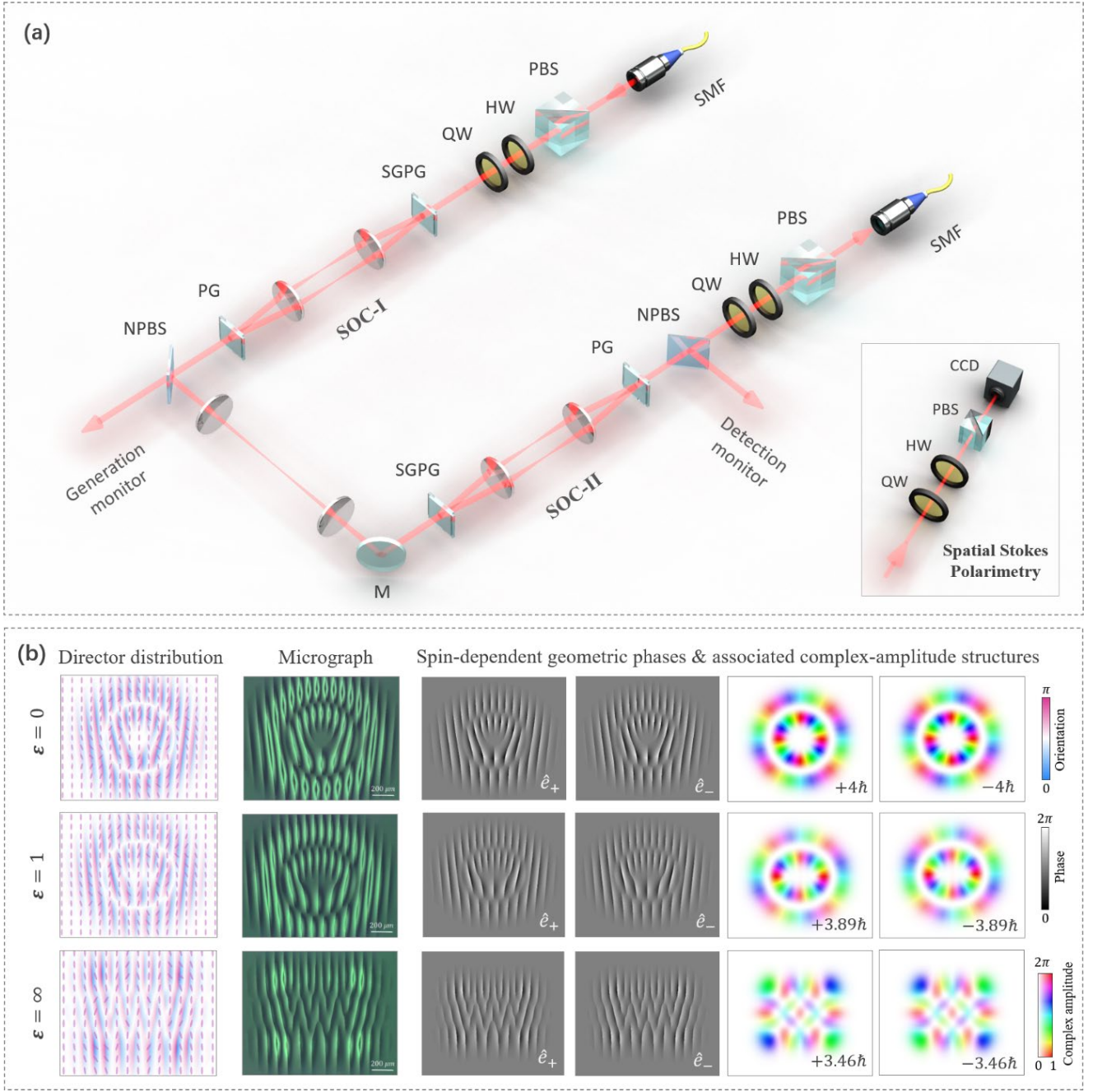

**Figure 2. (a)** Schematic representation of the experimental setup, where the key components are the single-mode fiber (SMF), half-wave plate (HW), quarter-wave plate (QW), polarizing beam splitter (PBS), nonpolarizing beam splitter (NPBS), structured geometric phase grating (SGPG), polarization grating (PG), and camera (CCD). The right-bottom inset shows the constitution of spatial Stokes polarimetry. **(b)** Schematics and characterizations of the SGPG design for generating $|\text{IG}_{6,4}^{\pm}, \hat{e}_{+}\rangle$ with $\varepsilon = 0, 1$ and $\infty$, including designed LC director distributions, polarizing micrographs taken at 0V voltage under crossed polarizers, and target geometric phases (GPs) & complex amplitudes.

## Experimental Results

Figure 2(a) shows the experimental setup used to verify the principle illustrated above. A narrow linewidth laser operating at 780 nm, collimated from a single-mode fiber as a TEM$_{00}$ beam, was used as the initial light source of the experiment. The TEM$_{00}$ beam was first adjusted to the desired polarization using a SoP-control unit consisting of a polarizing beam splitter (PBS) and two wave plates (QW and HW). This scalar polarized TEM$_{00}$ beam was then converted into a corresponding vectorial structured Gaussian beam by passing it through the first SOC convertor (SOC-I), that is, by performing SOC state generation. Then, the generated vector beam was guided into the second SOC convertor (SOC-II), which operated in the detection mode, and was converted back to a corresponding scalar polarized beam. All the LC elements were operated in the half-wave condition by controlling their load voltage (a 1 kHz square-wave signal with a peak-peak voltage of around 2 V is applied). To match the beam size and divergence in SOC state generation and detection, a 4f-system was placed between the two SOC convertors. Finally, another SoP-control unit was used to perform Stokes tomography on the scalar polarized beam output from SOC-II, and projective values (i.e., Stokes parameters) were received by a single-mode fiber. In addition, to an *in-situ* measure of the beam structure in SOC state generation and detection operations, we used spatial Stokes polarimetry, see the apparatus in the right-bottom inset in Fig. 2(a), to monitor polarizing beam patterns sampled from two nonpolarizing beam splitters (NPBSs).

In the experiment, to show the modal versatility of the principle (i.e., can generate and measure arbitrary spatial modes), SGPGs were designed to control a group of generalized SOC states in elliptical coordinates, i.e., vector Ince-Gauss (IG) modes [37]. As vector eigen solutions of the paraxial wave equation, they have a propagation-invariant beam structure but, unlike the CV mode, are usually not rotationally symmetric. This indicates that the helical mode pair $\psi_\pm(\boldsymbol{r},z)$ within the 'generalized' SOC state probably does not carry well-defined OAMs per photon. Specifically, here, helical IG modes $\mathrm{IG}_{6,4}^\pm = \sqrt{1/2}\,(\mathrm{IG}_{6,4}^e \pm i\mathrm{IG}_{6,4}^o)$ with three ellipticities $\varepsilon = 0, 1$ and $\infty$ (see Supporting Information) were chosen for $\psi_\pm(\boldsymbol{r},z)$, corresponding to a transition from $\mathrm{LG}_{\pm 4,1}$ to $\mathrm{IG}_{6,4}^\pm$ and finally to helical Hermite–Gauss (HG) modes $\sqrt{1/2}\,(\mathrm{HG}_{42} \pm i\mathrm{HG}_{33})$. Figure 2(b) shows the designed LC director distributions (left) and observed polarizing micrographs (middle) of the SGPGs used in experiments, as well as their spin-switchable geometric phase and corresponding complex amplitudes to be structured (right) in theory. In all three cases, we see that the structured complex amplitude conjugates have the same intensity profile but opposite OAMs, so each pair can form an SU(2) unit sphere regarding the OAM. In particular, the average OAMs per photon carried by the helical IG ($\pm 3.89\hbar$) and HG ($\pm 3.46\hbar$) modes were calculated by reconstructing them with superposition of LG modes with the same order $N = 2p + |\ell| = 6$ [54,55]. See Supporting Information for details on the paraxial Gaussian modes.

The experimental results are shown in Fig. 3. For each ellipticity, we chose four states, evenly spaced on the rainbow-colored circle on the higher-order Poincaré sphere, as desired SOC states to be generated and measured. The polarizing beam patterns shown in the first (Generation) column are experimentally generated vector modes that we observed using spatial Stokes polarimetry. As an example, black insets nearby the Generation column show the observed spatial Stokes projection of each state |a⟩. We see that the effect of the SOC state transition from state |a⟩ to |d⟩, or the motion on the higher-order Poincaré sphere, is reflected in the variation of SoP distributions. The position parameters, i.e., $a$ and $\theta$, according to Eq. (3), can be measured quantitatively by converting the vector mode back into the corresponding scalar one. The middle (Detection) column shows the observed polarizing beam patterns sampled from the NPBS after detection by SOC-II. The theory regarding the pattern evolution in vector mode detection is given in the Supporting Information. Here, only the central Gaussian-like patterns were coupled into the SMF, and we used SoPs covering them to determine the position parameters $a$ and $\theta$, as indicated by the colored dots embedded on the rainbow path to show the measured positions. For a more intuitive comparison, the right (Reconstruction) column shows the simulated polarizing beam pattern of each involved SOC state according to the measured position parameters. That is, these patterns were calculated using Eq. (1) with measured $a$ and $\theta$ as input. The theoretical reconstruction is shown to be highly consistent with that in the observation, verifying the precision of the device in both SOC state generation and detection.

## Conclusion & Discussion

We experimentally demonstrated a new type of photonic SOC device, the SGPG, using the LC geometric phase. This advanced SOC device, compared with the present generation of devices, such as the q-plate, enables conjugate complex-amplitude control of orthogonal circular polarizations by bringing in a spatially varied grating microstructure. Such a crucial advance unlocks the control of paraxial structured light in all spatial dimensions via geometric phase. Taking the results in Fig. 3(a) as an example, the accession of radial modes further boosts the dimensions (or data capacity) of a quantum (or spatial-division multiplex) system relative to those of a photonic SOC system controlled by q-plates. In addition to the special CV modes, as shown in Figs. 3(b) and (c), the device can control generalized SOC states without rotational symmetry (or well-defined OAM). This capability makes it a key extra-/intracavity component to build a structured laser that has greater tunability in beam structure, compared with reported systems based on q-plate and metasurface [56,57]. For quantum optics, the reciprocal SOC interface demonstrated here allows to implement a Bell measurement for arbitrary SOC states, which is the basis towards the teleportation scheme for SOC photon pairs [46]. Moreover, owing to the capability of full-field spatial mode control [58], the device also paves the way to quantum control of high-dimension photonic skyrmions [59,60].

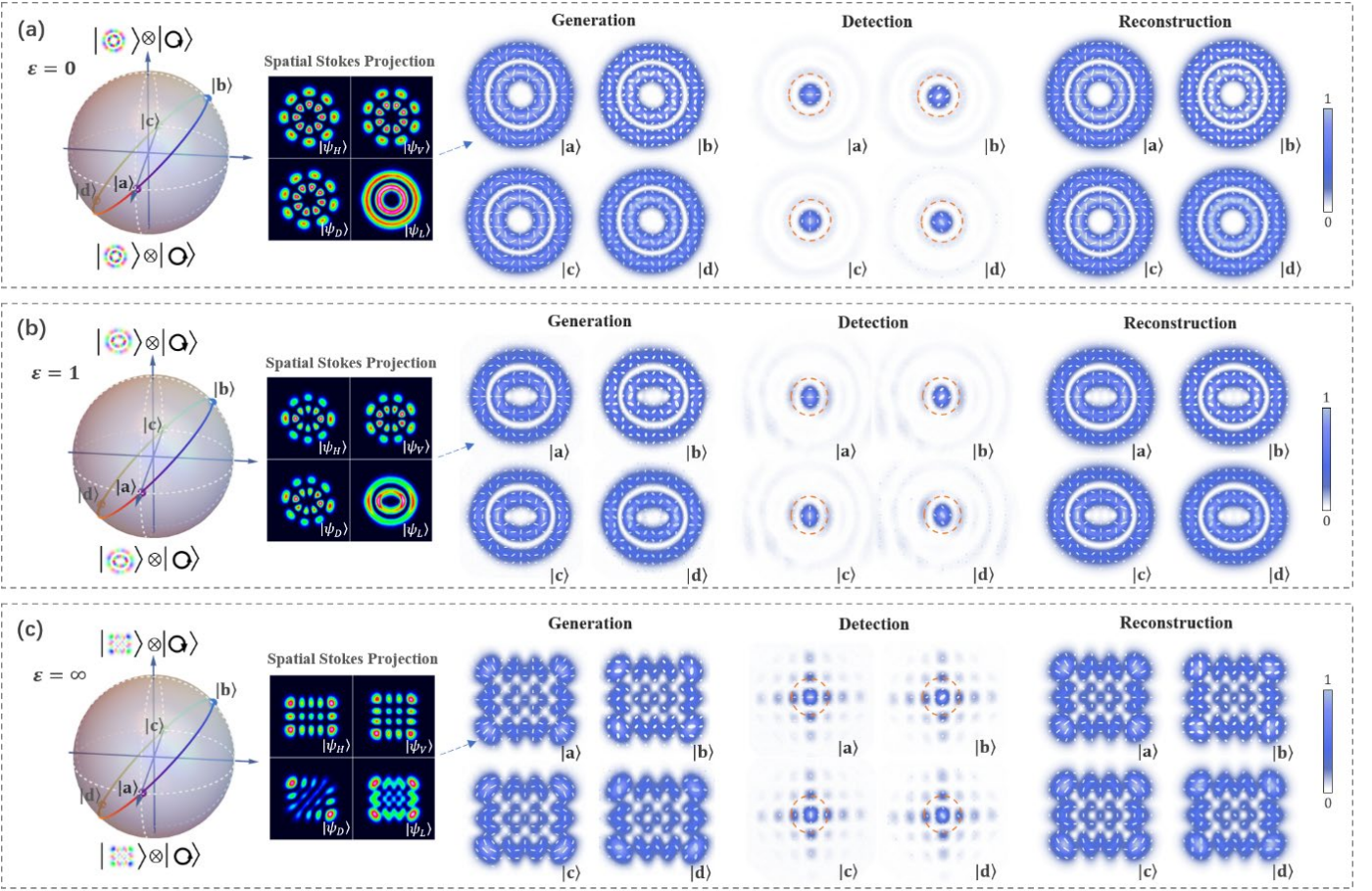

**Figure 3** Experimental results of the reciprocal SOC conversion used for the generation and detection of generalized SOC modes, where **(a)**–**(c)** correspond to vector IG modes with $\varepsilon = 0, 1$ and $\infty$, respectively. In the detection column, only the SoPs covering on the center patterns surrounded by orange dashed circles were coupled into the SMF and used to determine SOC states. Additional data acquired by spatial Stokes measurement are provided in the Supplementary Materials.

Beyond single-beam vector mode control, this principle can further realize multiple vector mode control through the addition of a Dammann grating structure [12, 61,62], see Supporting Information for an extended discussion on this. This represents a promising way to develop information exchange and processing units working for photonic SOC states, i.e., vector-mode multiplexers and demultiplexers. Moreover, the existence of the grating phase makes the diffraction response tunable with both the polarization and temporal spectrum of the incident beam, suggesting that the principle may also be used to shape and control the nonseparable structure of paraxial light in ray-wave coupled and spatiotemporal degrees of freedom [63-68]. Another issue of concern is the fabrication complexity, and another key advantage of SGPG is that it only requires the birefringence microstructure having a spatially varied orientation with global identical retardation, i.e., the complexity is same as the fabrication of q-plates. Hence, this new-type SOC device can be easily commercially manufactured, especially the LC version enabled by the photoalignment technique.

Regarding the limitation of the SOC device presented here, although the key LC element SGPG supports arbitrary complex-amplitude control with SoP switchable behaviour, the crucial amplitude control is realized by introducing spatially-varying grating; as a consequence, the generated or measured mode conjugates $\psi_{\pm}(\boldsymbol{r}, z)$ were separated to $\pm 1^{\text{st}}$-order diffractions. Such spatial separation of $\psi_{\pm}(\boldsymbol{r}, z)$ is useful for the applications that requires shaping arbitrary scalar mode with SoP switchable behaviour. For vector mode control, however, we have to use a 4*f*-PG system with the phase-locking capacity to recombine the spatially separated $\psi_{\pm}(\boldsymbol{r}, z)$, leading to the whole SOC device bulky. To further simplify the components of the device, in principle, the 4*f* lens can also be replaced by geometric phase elements. However, for single-beam vector mode control, reciprocal SOC conversion realized via a single flat element is preferred and thus is still a worthy issue to further research.

## Methods

**LC microstructure fabrication** — The sulfonic azo-dye SD1 (Dai-Nippon Ink and Chemicals, Japan) was dissolved in dimethylformamide at a concentration of 0.3 wt% and used as the photoalignment agent. Two indium-tin-oxide glasses were ultrasonic and UV-Ozone cleaned. After that, they were spin-coated with the SD1 solution at 800 rpm for 10 s and then 3000 rpm for 40 s. After curing at 100 °C for 10 min, the two glasses were assembled into a 6-μm-thick empty cell with the aid of spacers. Next, a digital-micromirror-device (DMD) based photoalignment system was used to fulfill the dynamic multi-step photoalignment process [69]. The DMD (Discovery 3000, Texas Instruments) consisted of 1024 × 768 micro-mirrors with the pixel size of 13.68 μm × 13.68 μm, and a 5× objective was used to further reduce the pixel pitch. When loading the designed alignment patterns, this system exposed the cell to UV light in a multi-step manner with a total dose of ≈ 5 J cm$^{-2}$. Each step corresponded to a specific exposure pattern and a certain UV linear polarization direction. Since SD1 molecules tend to reorient perpendicular to the UV polarization direction, the SD1 layers could be endowed with the desired alignment distribution after photoalignment process. Finally, the nematic LC E7 (HCCH, China) was filled into the cell at 70 °C via the capillary action. After cooling to room temperature, the LC molecules acquired the same director distribution as SD1 through intermolecular interactions.

## Acknowledgements

This work was supported by the National Natural Science Foundation of China (Grant Nos. 62075050, 62222507, 11934013, 61975047, 12004175, and 62175101), the Innovation Program for Quantum Science and Technology (No. 2021ZD0301500), and the Natural Science Foundation of Jiangsu Province (Nos. BK20212004, and BK20200311).

## Reference List

[1] H. Rubinsztein-Dunlop, A. Forbes, M. V. Berry, et al. Roadmap on structured light. Journal of Optics. **19**, 013001 (2016). https://doi.org/10.1088/2040-8978/19/1/013001

[2] J. Chen, C. Wan and Q. Zhan. Vectorial optical fields: recent advances and future prospects. Science Bulletin. **63**, 54-74 (2018). https://doi.org/10.1016/j.scib.2017.12.014

[3] J. Wang, F. Castellucci and S. Franke-Arnold. Vectorial light–matter interaction: Exploring spatially structured complex light fields. AVS Quantum Science. **2**, 031702 (2020). https://doi.org/10.1116/5.0016007

[4] Forbes, M. De Oliveira and M. R. Dennis. Structured light. Nature Photonics. **15**, 253-262 (2021). https://doi.org/10.1038/s41566-021-00780-4

[5] D. Pohl. Operation of a ruby laser in the purely transverse electric mode TE01. Applied Physics Letters. **20**, 266-267 (1972). https://doi.org/10.1063/1.1654142

[6] Allen L, Beijersbergen M, Spreeuw R, et al. Orbital angular momentum of light and the transformation of Laguerre-Gaussian laser modes. Physical Review A. **45**, 8185-8189 (1992). https://doi.org/10.1103/PhysRevA.45.8185

[7] Y. Shen, X. Wang, Z. Xie, et al. Optical vortices 30 years on: OAM manipulation from topological charge to multiple singularities. Light: Science Applications. **8**, 1-29 (2019). https://doi.org/10.1038/s41377-019-0194-2

[8] K. Y. Bliokh, F. J. Rodríguez-Fortuño, F. Nori, et al. Spin–orbit interactions of light. Nature Photonics. **9**, 796-808 (2015). https://doi.org/10.1038/nphoton.2015.201

[9] V. Liberman and B. Y. Zel'dovich. Spin-orbit interaction of a photon in an inhomogeneous medium. Physical Review A. **46**, 5199 (1992). https://doi.org/10.1103/PhysRevA.46.5199

[10] Borges, M. Hor-Meyll, J. Huguenin, et al. Bell-like inequality for the spin-orbit separability of a laser beam. Physical Review A. **82**, 033833 (2010). https://doi.org/10.1103/PhysRevA.82.033833

[11] Rosales-Guzman, B. Ndagano and A. Forbes. A review of complex vector light fields and their applications. Journal of Optics. **20**, 31 (2018). https://doi.org/10.1088/2040-8986/aaeb7d

[12] Rosales-Guzmán and A. Forbes. *How to shape light with spatial light modulators*. (SPIE Press, 2017).

[13] Cardano and L. Marrucci. Spin–orbit photonics. Nature Photonics. **9**, 776-778 (2015). https://doi.org/10.1038/nphoton.2015.232

[14] H. Dorrah and F. Capasso. Tunable structured light with flat optics. Science. **376**, eabi6860 (2022). https://doi.org/10.1126/science.abi6860

[15] P. Chen, B. Y. Wei, W. Hu, et al. Liquid-Crystal-Mediated Geometric Phase: From Transmissive to Broadband Reflective Planar Optics. Advanced Materials. **32**, 21 (2020). https://doi.org/10.1002/adma.201903665

[16] C. Cisowski, J. Götte and S. Franke-Arnold. Colloquium: Geometric phases of light: Insights from fiber bundle theory. Reviews of Modern Physics. **94**, 031001 (2022). https://doi.org/10.48550/arXiv.2202.04356

[17] Cohen, H. Larocque, F. Bouchard, et al. Geometric phase from Aharonov–Bohm to Pancharatnam–Berry and beyond. Nature Reviews Physics. **1**, 437-449 (2019). https://doi.org/10.1038/s42254-019-0071-1

[18] S. Pancharatnam, Generalized theory of interference and its applications, 1956-Proceedings of the Indian Academy of Sciences-Section A.

[19] R. C. Devlin, A. Ambrosio, N. A. Rubin, et al. Arbitrary spin-to–orbital angular momentum conversion of light. Science. **358**, 896-901 (2017). https://doi.org/10.1126/science.aao5392

[20] L. Marrucci, C. Manzo and D. Paparo. Optical spin-to-orbital angular momentum conversion in inhomogeneous anisotropic media. Physical review letters. **96**, 163905 (2006). https://doi.org/10.1103/PhysRevLett.96.163905

[21] Q. W. Zhan. Cylindrical vector beams: from mathematical concepts to applications. Advances in Optics and Photonics. **1**, 1-57 (2009). https://doi.org/10.1364/aop.1.000001

[22] E. Nagali, L. Sansoni, L. Marrucci, et al. Experimental generation and characterization of single-photon hybrid ququarts based on polarization and orbital angular momentum encoding. Physical Review A. **81**, 052317 (2010). https://doi.org/10.1103/PhysRevA.81.052317

[23] C. Souza, C. Borges, A. Khoury, et al. Quantum key distribution without a shared reference frame. Physical Review A. **77**, 032345 (2008). https://doi.org/10.1080/09500340.2018.1455913

[24] R. Dorn, S. Quabis and G. Leuchs. Sharper Focus for a Radially Polarized Light Beam. Physical Review Letters. **91**, 233901 (2003). https://doi.org/10.1103/PhysRevLett.91.233901

[25] Vallone, V. D'ambrosio, A. Sponselli, et al. Free-space quantum key distribution by rotation-invariant twisted photons. Physical review letters. **113**, 060503 (2014). https://doi.org/10.1103/PhysRevLett.113.060503

[26] Sit, F. Bouchard, R. Fickler, et al. High-dimensional intracity

quantum cryptography with structured photons. Optica. **4**, 1006-1010 (2017). https://doi.org/10.1364/OPTICA.4.001006

[27] Ndagano, B. Perez-Garcia, F. S. Roux, et al. Characterizing quantum channels with non-separable states of classical light. Nature Physics. **13**, 397-402 (2017). https://doi.org/10.1038/nphys4003

[28] V. D'ambrosio, N. Spagnolo, L. Del Re, et al. Photonic polarization gears for ultra-sensitive angular measurements. Nature communications. **4**, 1-8 (2013). https://doi.org/10.1038/ncomms3432

[29] E. Karimi, G. Zito, B. Piccirillo, et al. Hypergeometric-gaussian modes. Opt. Lett. **32**, 3053-3055 (2007). https://doi.org/10.1364/ol.32.003053

[30] Z. Y. Zhou, Z. H. Zhu, S. L. Liu, et al. Quantum twisted double-slits experiments: confirming wavefunctions' physical reality. Science bulletin. **62**, 1185-1192 (2017). https://doi.org/10.1016/j.scib.2017.08.024

[31] E. Karimi, R. Boyd, P. De La Hoz, et al. Radial quantum number of Laguerre-Gauss modes. Physical review A. **89**, 063813 (2014). https://doi.org/10.1103/PhysRevA.89.063813

[32] E. Karimi, D. Giovannini, E. Bolduc, et al. Exploring the quantum nature of the radial degree of freedom of a photon via Hong-Ou-Mandel interference. Physical Review A. **89**, 013829 (2014). https://doi.org/10.1103/PhysRevA.89.013829

[33] Y. Zhang, F. S. Roux, M. Mclaren, et al. Radial modal dependence of the azimuthal spectrum after parametric down-conversion. Physical Review A. **89**, 043820 (2014). https://doi.org/10.1103/PhysRevA.89.043820

[34] V. Salakhutdinov, E. Eliel and W. Löffler. Full-field quantum correlations of spatially entangled photons. Physical review letters. **108**, 173604 (2012). https://doi.org/10.1103/PhysRevLett.108.173604

[35] L. Chen, T. Ma, X. Qiu, et al. Realization of the Einstein-Podolsky-Rosen paradox using radial position and radial momentum variables. Physical review letters. **123**, 060403 (2019). https://doi.org/10.1103/PhysRevLett.123.060403

[36] R. Y. Zhong, Z. H. Zhu, H. J. Wu, et al. Gouy-phase-mediated propagation variations and revivals of transverse structure in vectorially structured light. Physical Review A. **103**, 9 (2021). https://doi.org/10.1103/PhysRevA.103.053520

[37] Yao-Li, X.-B. Hu, B. Perez-Garcia, et al. Classically entangled Ince–Gaussian modes. Applied Physics Letters. **116**, 221105 (2020). https://doi.org/10.1063/5.0011142

[38] M. A. Bandres and J. C. Gutiérrez-Vega. Ince–gaussian beams. Opt. Lett. **29**, 144-146 (2004). https://doi.org/10.1364/OL.29.000144

[39] F. Brandt, M. Hiekkamäki, F. Bouchard, et al. High-dimensional quantum gates using full-field spatial modes of photons. Optica. **7**, 98-107 (2020). https://doi.org/10.1364/OPTICA.375875

[40] J. Wu, B. S. Yu, Z. H. Zhu, et al. Conformal frequency conversion for arbitrary vectorial structured light. Optica. **9**, 187-196 (2022). https://doi.org/10.48550/arXiv.2109.13636

[41] D. Sugic, R. Droop, E. Otte, et al. Particle-like topologies in light. Nature communications. **12**, 1-10 (2021). https://doi.org/10.1038/s41467-021-26171-5

[42] B. P. Da Silva, V. Pinillos, D. Tasca, et al. Pattern revivals from fractional Gouy phases in structured light. Physical review letters. **124**, 033902 (2020). https://doi.org/10.1103/PhysRevLett.124.033902

[43] L. Pereira, A. Khoury and K. Dechoum. Quantum and classical separability of spin-orbit laser modes. Physical Review A. **90**, 053842 (2014). https://doi.org/10.1103/PhysRevA.90.053842

[44] G. Milione, H. Sztul, D. Nolan, et al. Higher-order Poincaré sphere, Stokes parameters, and the angular momentum of light. Physical review letters. **107**, 053601 (2011). https://doi.org/10.1103/PhysRevLett.107.053601

[45] R. Gutiérrez-Cuevas, S. Wadood, A. Vamivakas, et al. Modal Majorana sphere and hidden symmetries of structured-Gaussian beams. Physical Review Letters. **125**, 123903 (2020). https://doi.org/10.1103/PhysRevLett.125.123903

[46] Z. Khoury and P. Milman. Quantum teleportation in the spin-orbit variables of photon pairs. Physical Review A. **83**, 060301 (2011). https://doi.org/10.1103/PHYSREVA.83.060301

[47] P. Chen, L. L. Ma, W. Duan, et al. Digitalizing self-assembled chiral superstructures for optical vortex processing. Advanced Materials. **30**, 1705865 (2018). https://doi.org/10.1002/adma.201705865

[48] P. Chen, Z. X. Shen, C. T. Xu, et al. Simultaneous Realization of Dynamic and Hybrid Multiplexed Holography via Light-Activated Chiral Superstructures. Laser Photonics Reviews. 2200011 (2022). https://doi.org/10.1002/lpor.202200011

[49] P. Chen, L.-L. Ma, W. Hu, et al. Chirality invertible superstructure mediated active planar optics. Nature communications. **10**, 1-7 (2019). https://doi.org/10.1038/s41467-019-10538-w

[50] L. Zhu, C. T. Xu, P. Chen, et al. Pancharatnam–Berry phase reversal via opposite-chirality-coexisted superstructures. Light: Science&Applications. **11**, 1-8 (2022). https://doi.org/10.1038/s41377-022-00835-3

[51] E. Toninelli, B. Ndagano, A. Vallés, et al. Concepts in quantum state tomography and classical implementation with intense light: a tutorial. Advances in Optics Photonics. **11**, 67-134 (2019). https://doi.org/10.1364/AOP.11.000067

[52] B. Ndagano, I. Nape, M. A. Cox, et al. Creation and detection of vector vortex modes for classical and quantum communication. Journal of Lightwave Technology. **36**, 292-301 (2018). https://doi.org/10.1109/JLT.2017.2766760

[53] B. S. Yu, C. Y. Li, Y. Yang, et al. Directly Determining Orbital Angular Momentum of Ultrashort Laguerre–Gauss Pulses via Spatially-Resolved Autocorrelation Measurement. Laser Photonics Reviews. 2200260 (2022). https://doi.org/10.1002/lpor.202200260

[54] R. Yang, H. J. Wu, W. Gao, et al. Parametric upconversion of Ince-Gaussian modes. Opt. Lett. **45**, 3034-3037 (2020). https://doi.org/10.1364/ol.393146

[55] W. N. Plick, M. Krenn, R. Fickler, et al. Quantum orbital angular momentum of elliptically symmetric light. Physical Review A. **87**, 033806 (2013). https://doi.org/10.1103/PhysRevA.87.033806

[56] H. Sroor, Y.-W. Huang, B. Sephton, et al. High-purity orbital angular momentum states from a visible metasurface laser. Nature Photonics. **14**, 498-503 (2020). https://doi.org/10.1038/s41566-020-0623-z

[57] D. Naidoo, F. S. Roux, A. Dudley, et al. Controlled generation of higher-order Poincaré sphere beams from a laser. Nature Photonics. **10**, 327-332 (2016). https://doi.org/10.1038/nphoton.2016.37

[58] F. Brandt, M. Hiekkamäki, F. Bouchard, et al. High-dimensional quantum gates using full-field spatial modes of photons. Optica. **7**, 98-107 (2020). https://doi.org/10.1364/OPTICA.375875

[59] Y. Shen, B. Yu, H. Wu, et al. Topological transformation and free-space transport of photonic hopfions. Advanced Photonics. **8**, 1-29 (2019). https://doi.org/10.1117/1.AP.5.1.015001

[60] D. Sugic, R. Droop, E. Otte, et al. Particle-like topologies in light. Nature communications. **12**, 6785 (2021). https://doi.org/10.1038/s41467-021-26171-5

[61] P. Chen, S. J. Ge, W. Duan, et al. Digitalized Geometric Phases for Parallel Optical Spin and Orbital Angular Momentum Encoding. ACS Photonics. **4**, 1333-1338 (2017). https://doi.org/10.1021/ACSPHOTONICS.7B00263

[62] S. Liu, P. Chen, S. Ge, et al. 3D Engineering of Orbital Angular Momentum Beams via Liquid-Crystal Geometric Phase. Laser Photonics Reviews. **16**, 2200118 (2022). https://doi.org/10.1002/lpor.202200118

[63] Y. Shen and C. Rosales-Guzmán. Nonseparable states of light: From quantum to classical. Laser Photonics Reviews. 2100533 (2022). https://doi.org/10.48550/arXiv.2203.00994

[64] C. He, Y. Shen and A. Forbes. Towards higher-dimensional structured light. Light: Science Applications. **11**, 1-17 (2022). https://doi.org/10.1038/s41377-022-00897-3

[65] Y. Shen, X. Yang, D. Naidoo, et al. Structured ray-wave vector vortex beams in multiple degrees of freedom from a laser. Optica. **7**, (2020). https://doi.org/10.1364/optica.382994

[66] Y. Shen, I. Nape, X. Yang, et al. Creation and control of high-dimensional multi-partite classically entangled light. Optica. **10**, 10 (2021). https://doi.org/10.1038/s41377-021-00493-x

[67] C. Wan, Q. Cao, J. Chen, et al. Toroidal vortices of light. Nature Photonics. **16**, 519-522 (2022). https://doi.org/10.1038/s41566-022-01013-y

[68] C. Wan, J. Chen, A. Chong, et al. Photonic orbital angular momentum with controllable orientation. National Science Review. **9(7)**, 149 (2022). https://doi.org/10.1093/nsr/nwab149

[69] Ji, W., Lee, CH., Chen, P. et al. Meta-q-plate for complex beam shaping. Sci Rep 6, 25528 (2016). https://doi.org/10.1038/srep25528